\documentclass[12pt,a4paper]{article}
\usepackage{amsmath, amssymb}
\usepackage{amsfonts}
\usepackage[mathscr]{eucal}
\usepackage{ascmac}
\usepackage[dvipdfmx]{graphicx}
\usepackage{subfigure}
\usepackage[charter]{mathdesign}
\usepackage{bm}
\usepackage{cite}
\usepackage{here}
\usepackage{color}

\usepackage[height=22.5cm,width=16.5cm]{geometry}
\begin{document}
\renewcommand{\thefootnote}{$\clubsuit$\arabic{footnote}}
\def\a{\alpha}
\def\b{\beta}
\def\c{\varepsilon}
\def\d{\delta}
\def\e{\epsilon}
\def\f{\phi}
\def\g{\gamma}
\def\h{\theta}
\def\k{\kappa}
\def\l{\lambda}
\def\m{\mu}
\def\n{\nu}
\def\p{\psi}
\def\q{\partial}
\def\r{\rho}
\def\s{\sigma}
\def\t{\tau}
\def\u{\upsilon}
\def\v{\varphi}
\def\w{\omega}
\def\x{\xi}
\def\y{\eta}
\def\z{\zeta}
\def\D{\Delta}
\def\G{\Gamma}
\def\H{\Theta}
\def\L{\Lambda}
\def\F{\Phi}
\def\P{\Psi}
\def\S{\Sigma}

\def\der{\partial}
\def\o{\over}
\def\beq{\begin{align}}
\def\eeq{\end{align}}
\newcommand{\gsim}{ \mathop{}_{\textstyle \sim}^{\textstyle >} }
\newcommand{\lsim}{ \mathop{}_{\textstyle \sim}^{\textstyle <} }
\newcommand{\vev}[1]{ \left\langle {#1} \right\rangle }
\newcommand{\bra}[1]{ \langle {#1} | }
\newcommand{\ket}[1]{ | {#1} \rangle }
\newcommand{\EV}{ {\rm eV} }
\newcommand{\KEV}{ {\rm keV} }
\newcommand{\MEV}{ {\rm MeV} }
\newcommand{\GEV}{ {\rm GeV} }
\newcommand{\TEV}{ {\rm TeV} }
\def\diag{\mathop{\rm diag}\nolimits}
\def\Spin{\mathop{\rm Spin}}
\def\SO{\mathop{\rm SO}}
\def\O{\mathop{\rm O}}
\def\SU{\mathop{\rm SU}}
\def\U{\mathop{\rm U}}
\def\Sp{\mathop{\rm Sp}}
\def\SL{\mathop{\rm SL}}
\def\tr{\mathop{\rm tr}}
\def\Dterm{_{\theta^2\bar{\theta^2}}}
\def\Fterm{_{\theta^2}}
\def\F*term{_{\bar{\theta}^2}}
\makeatletter
\@addtoreset{equation}{section}
\def\theequation{\thesection.\arabic{equation}}
\makeatother
\def\mi{m_{\phi}}
\def\mpl{M_{\rm pl}}

\def\IJMP{Int.~J.~Mod.~Phys. }
\def\MPL{Mod.~Phys.~Lett. }
\def\NP{Nucl.~Phys. }
\def\PL{Phys.~Lett. }
\def\PR{Phys.~Rev. }
\def\PRL{Phys.~Rev.~Lett. }
\def\PTP{Prog.~Theor.~Phys. }
\def\ZP{Z.~Phys. }


\newcommand{\TODO}[1]{{$[[ \clubsuit\clubsuit$ \bf #1 $\clubsuit\clubsuit ]]$}}
\newcommand{\kmrem}[1]{{\color{red} \bf $[[ $ KM: #1$ ]]$}}

\begin{titlepage}

\begin{flushright}
IPMU 13-0233\\
UT-13-43
\end{flushright}

\vskip 3cm

\begin{center}
{\LARGE \bfseries
Thermalization after/during Reheating\\
}

\vskip .65in
{\large
Keisuke Harigaya$^{\diamondsuit}$ and Kyohei Mukaida$^{\spadesuit}$
}

\vskip .4in
\begin{tabular}{ll}
$^{\diamondsuit}$ &\!\! {\em Kavli Institute for the Physics and
 Mathematics of the Universe (WPI), }\\
&{\em University of Tokyo,  Kashiwa, Chiba 277-8583, Japan}\\[.5em]
$^{\spadesuit}$ & \!\! {\em Department of Physics, Faculty of Science, }\\
& {\em University of Tokyo,  Bunkyo-ku, Tokyo 133-0033, Japan}
\end{tabular}

\vskip .75in

\begin{abstract}
If reheating of the Universe takes place via Planck-suppressed decay,
it seems that the thermalization of produced particles might be delayed,
since they have large energy/small number densities
and number violating large angle scatterings which decrease the momentum of particles
by large amount
are inefficient correspondingly.
In this paper, we study the thermalization of such ``under occupied'' decay products in detail,
following recent developments in understanding the thermalization of non-abelian plasma.
Contrary to the above naive expectation,
it is shown that in most cases thermalization after/during reheating 
occurs instantaneously 
by properly taking account of scatterings with small angles and of particles with small momenta.
In particular, the condition for instantaneous thermalization before the completion of reheating is found to be
$\alpha^{8/5} \gg (m_\phi / \mpl) (\mpl^2 \Gamma_\phi / m_\phi^3)^{1/5}$,
which is much milder than that obtained in previous works 
with small angle scatterings taken into account.
\end{abstract}

\end{center}

\end{titlepage}

\newpage

\tableofcontents

\section{Introduction}
\label{sec:intro}
Thermodynamics plays a central role in the standard cosmology.
The early Universe is much denser than the present one and hence
is filled with fluid of plasma, whose dynamics is
described by thermodynamics and simply characterized by a temperature.
This picture is strongly supported by
the observation of the cosmic microwave background~\cite{Penzias:1965wn,Mather:1993ij}
and the success of the big-bang nucleosynthesis (BBN) \cite{Schramm:1997vs}.

Tracing back the history of the Universe to the past, we encounter an era
in which the energy density of the Universe is dominated by a potential
energy of a scalar field, such as an inflaton~\cite{Guth:1980zm}, a curvaton~\cite{Linde:1996gt}, 
and a moduli field~\cite{Coughlan:1983ci},
and hence the Universe is expected to be far from thermal equilibrium.
The scalar field should eventually decay and yield its energy to
radiation. This process is refered to as ``reheating''.

Here, a fundamental question arises. 
Does radiation thermalize soon
after reheating?\footnote{
	Throughout this paper, ``thermalization'' means that the momentum distribution of particles
	are given by chemical equilibrium ones.
}
This consideration is necessary in order to 
determine whether or not radiation is simply characterized by the temperature
and hence is practically important for particle cosmology.\footnote{
	Otherwise, the property of radiation strongly depends on its initial conditions.
}
For example, the thermal leptogenesis~\cite{Fukugita:1986hr} requires a
temperature larger than $10^9$ GeV~\cite{Buchmuller:2004nz},
the gravitino abundance depends on the temperature of the Universe~\cite{Kawasaki:1994af},
and the efficiency of the Affleck-Dine mechanism~\cite{Affleck:1984fy} might depend on 
the property of thermal plasma~\cite{Dine:1995kz,Anisimov:2000wx}.

Thermalization of radiation after reheating has been investigated in
the literature.
It is sometimes argued that thermalization is reached well after
reheating in certain cases when the decay of the scalar field
is small,\footnote{
	A small decay width is expected in order to suppress quantum corrections 
	to the inflaton potential~\cite{Chung:1998rq}.
}
for example if it is induced by Planck-suppressed interactions.
In this case,
the produced plasma is energetic but has a small number density
compared with thermal one,
and hence inelastic number violating processes play crucial roles.
In the literature~\cite{Ellis:1987rw,McDonald:1999hd,Allahverdi:2000ss,Mazumdar:2013gya}, 
thermalization is assumed to occur by
accumulations of large angle scatterings, or, hard processes.
Then, their interaction rates are suppressed by the large energy scale,
which is around the mass of the scalar field (\textit{e.g.}, inflaton),
and as a result it is concluded that thermalization might be delayed.

However,
there is another limit: number violating scatterings with small angles.
Then, their interaction rates are enhanced though energy loss per one event is suppressed.
In Refs.~\cite{Chung:1998rq,Davidson:2000er}, it is suggested that the enhanced small angle scatterings 
play important roles in thermalization.
It is concluded that thermalization is completed instantaneously 
after reheating as long as the mass of the scalar field is smaller than $10^{13}\,\GEV$~\cite{Davidson:2000er}.

At the same time,
in the context of the quark gluon plasma (QGP),
more thorough study is performed with hot QCD theories
while paying attentions to effects of small angle
scatterings and to the evolution of particles with small momenta, or, soft processes~\cite{Baier:2000sb,Kurkela:2011ti}. 
Importantly, it is pointed out that thermalization is dominated by soft processes in the case of sparse number density,
and thermalization proceeds from a soft sector, that is, so-called ``bottom-up thermalization'' 
takes place~\cite{Baier:2000sb}.

In this paper, we follow and generalize the discussion in Ref.~\cite{Kurkela:2011ti},
and estimate thermalization time scales by gauge interactions.
We consider a thermalization
process after/during reheating
while paying attention to the red-shift of momenta and the dilution of the number density by a cosmic expansion,
and provide with conditions under
which radiation thermalizes soon after they are emitted from the scalar field.
We show that thermalization after/during reheating occurs instantaneously in most cases
within the cosmic time scale
by properly taking small angle scatterings and the evolution of the soft sector into account.
In particular, it is found that the condition for instantaneous thermalization,
$\alpha^{8/5} \gg (m_\phi / \mpl) (\mpl^2 \Gamma_\phi / m_\phi^3)^{1/5}$,
is much milder than that obtained in Ref.~\cite{Davidson:2000er},
since the growth of the soft sector promotes the energy dissipation of decay products
with large momenta.
As a result, it is found that thermalization is completed instantaneously
even if the mass of the scalar field is as large as $10^{15}\,\GEV$,
which is the case with inflation models discussed in Refs.~\cite{Dimopoulos:1997fv,Takahashi:2010ky,Harigaya:2012pg}.

This paper is organized as follow.
In Sec.~\ref{sec:review}, we briefly review how reheating proceeds.
In Sec.~\ref{sec:naive}, a thermalization process solely by large
angle scatterings is considered.
In Sec.~\ref{sec:under}, we discuss how thermalization proceeds and estimate
thermalization time scales by taking soft processes into account,
following Ref.~\cite{Kurkela:2011ti}.
Then, thermalization after/during reheating is studied in
Sec.~\ref{sec:application}.
The last section is devoted to conclusion and discussions.

\section{Brief review of the reheating process}
\label{sec:review}
In this section, we review a reheating process in which an oscillating scalar
field $\phi(t)$ loses its energy through a friction by the expansion of the Universe
(the Hubble friction) and decays into lighter particles.
Throughout this paper, we concentrate on the case in which the scalar field dominates the
energy density of the Universe and the decay is well described by a
perturbative decay.%
\footnote{
If the oscillation time scale of the scalar field is much slower than the typical interaction time scale of background plasma,
the interactions of the decay products of the scalar field with the
radiation affect the dissipation process~\cite{Yokoyama:2004pf,BasteroGil:2010pb,Drewes:2010pf,Mukaida:2012qn},
and the discussion on the
thermalization process strongly depends on the detail of the theory.
}

In the expanding Universe,
the energy density of the oscillating scalar field $\rho_\phi$ and that of the
decay products $\rho_r$ follow the equations of motion, 
\begin{align}
\frac{{\rm d}}{{\rm d}t} \rho_\phi +\left(3H +
\label{eq:energy_scalar}			
	     \Gamma_\phi\right)\rho_\phi=0,\\
\label{eq:energy_radiation}
 \frac{{\rm d}}{{\rm d}t}\rho_r +4H\rho_r - \Gamma_\phi \rho_\phi=0,\\
\label{eq:Hubble}
3H^2 \mpl^2 = \rho_\phi + \rho_r,
\end{align}
where $H$ and $\Gamma_\phi$ are
the Hubble parameter of the Universe and the decay rate of the scalar field,
respectively. Here, we have assumed that the self interaction of the
scalar field is negligible, which is a good approximation in the last
stage of the oscillation. We have also assumed that the decay products
are relativistic particles.

At the early stage of the oscillation, the energy density of the
Universe is dominated by the scalar 
field and hence the Hubble parameter is
given by $H=\sqrt{\rho_\phi}/\left(\sqrt{3}\mpl\right)$.
Also, the decay width is negligible in comparison with the Hubble parameter.
Within this approximation, the solution of
Eq. (\ref{eq:energy_scalar}-\ref{eq:Hubble}) is given by
\begin{align}
 \rho_\phi \simeq \frac{4}{3}\frac{\mpl^2}{t^2},~~~
\rho_r \simeq  \frac{4}{5}\Gamma_\phi \mpl^2 \frac{1}{t}\sim \rho_\phi
\Gamma_\phi t,~~~
H=\frac{2}{3t}.
\end{align}

The energy density of the radiation is dominated by particles with momentum $p\sim m_\phi$, and hence the distribution function at that momentum is given by
\begin{align}
 f(p\sim m_\phi)\sim \frac{\rho_r}{m_{\phi}^4} \sim \frac{\Gamma_\phi \mpl^2}{m_\phi^4 t},
\end{align}
for each time.
Note that the scalar field produces particles with $p\sim m_{\phi}$ and particles with lower momenta are provided by the red-shift.
Since the momentum is red-shifted in proportion to $t^{-2/3}$ during a matter dominated era, the distribution function 
is given by~\cite{Allahverdi:2000ss}
\begin{align}
	f(p) \sim 
	\begin{cases}
		\left( \cfrac{\Gamma_\phi \mpl^2}{m_\phi^3} \right) \left( m_\phi t \right)^{-1}
		\left( \cfrac{m_\phi}{p} \right)^{3/2} &\text{for}~~(t/t_i)^{-2/3} m_\phi \lesssim p \lesssim m_\phi \\[10pt]
		0 &\text{for~~otherwise}
	\end{cases},
\label{eq:dist_dr_reh_1}
\end{align}
if the produced particles do not interact at all.
Here, $t_i$ is the time at which the oscillation of the scalar field begins to dominate the energy density of the Universe.\footnote{
Below the infrared edge, there is a contribution from the decay of the scalar field before its oscillation dominates the energy density of the Universe in general. However, the shape of the contribution is always shallower than $p^{-3/2}$. Further, as we will see later, the time scale of thermalization is governed by the distribution of the hardest particles. Therefore, we neglect such infrared contributions in the following.
}

The energy densities of the scalar field and of the radiation become
comparable at $t^{-1}\sim H\sim \Gamma_\phi$.
After this point, the Universe is radiation-dominant.
This point is referred to as ``the end of reheating'', and the
``reheating temperature'' $T_\text{rh}$ is often defined by
\begin{align}
 3\Gamma_\phi^2 \mpl^2 \equiv \frac{\pi^2}{30}g_* T_\text{rh}^4,
\end{align}
where $g_*$ is the effective degree of freedom of would-be thermalized particles.
If the decay products of the scalar field interact with each others
sufficiently fast, they thermalize soon after the end of
reheating. $T_\text{rh}$ coincides with the temperature of the Universe at that point.
In the following sections, we discuss how thermalization proceeds by interactions among decay products.

\section{First look at thermalization processes: effects of large angle scatterings}
\label{sec:naive}

In this section, we consider a thermalization process solely by large angle
scatterings, as is assumed in Refs.~\cite{Ellis:1987rw,McDonald:1999hd,Allahverdi:2000ss,Mazumdar:2013gya}.
Here, ``large angle'' means that the scatterings involve large momentum
exchanges so that particles lose their energy efficiently.
We will see that large angle scatterings cannot bring about
instantaneous thermalization after reheating, if the decay
rate of the scalar field is small, such as the one induced by Planck-suppressed interactions.
The result shows the necessity of consideration of thermalization
processes with cares.

In order to investigate whether thermalization occurs soon after the
end of reheating,
let us first estimate the number density of the decay products.
Just after the end of reheating, the number density of
the decay products, $n_i$,
is as large as
\begin{align}
 n_i \simeq \left. \frac{\rho_\phi}{m_{\phi}} \right|_{t\sim \Gamma_\phi^{-1}} \simeq \frac{\Gamma_\phi^2 \mpl^2}{m_\phi},
\end{align}
where we have used the result obtained in the previous section.
As we have mentioned in the previous section, the momentum of decay
products, $p_i$, is as large as
the mass of the scalar field, $p_i\simeq m_{\phi}$.

Let us compare this number density with a thermal one.
If thermalization occurs instantaneously, number densities of
particles, $n_{\rm th}$, are as large as
\begin{align}
n_{\rm th} \simeq T_\text{rh}^3 \simeq \Gamma_\phi^{3/2}\mpl^{3/2}.
\end{align}
Its ratio to $n_i$ is given by
\begin{align}
 \frac{n_{\rm th}}{n_i}\simeq \frac{m_{\phi}}{\sqrt{\Gamma_\phi \mpl}}.
\end{align}
For example, let us consider two extreme cases, in which the decay occurs
due to an order one Yukawa coupling and a dimension $5$ Planck-suppressed interaction.
Decay rates are as large as
\begin{align}
 \Gamma_{\phi} \simeq
\left\{
\begin{array}{ll}
 m_{\phi}  & (\text{Yukawa})\\
 m_{\phi}^3/\mpl^2& (\text{dimension $5$ Planck-suppressed})
\end{array}
\right..
\end{align}
Accordingly, the ratio of the number densities is given by
\begin{align}
 n_{\rm th}/n_i \simeq
\left\{
\begin{array}{ll}
 \sqrt{m_\phi/\mpl}\ll1  & (\text{Yukawa})\\
 \sqrt{\mpl/m_{\phi}} \gg1& (\text{dimension $5$ Planck-suppressed})
\end{array}
\right..
\end{align}
Therefore, if the decay occurs via the order one Yukawa coupling, the
number density is larger than the thermalized one, in
other words, the Universe is over occupied.
On the other hand, if the decay occurs via the Planck-suppressed interaction, the
number density is smaller than the thermalized one, in
other words, the Universe is under occupied.

For the over occupied case, $T_\text{rh}$ is larger than
$m_{\phi}$.
In such a case, it is known that thermal effects play important role in the
reheating process~\cite{Mukaida:2012bz,Drewes:2013iaa}, and the resulting tempetarure of the Universe 
depends on the detail of the theory.\footnote{
	For some cases,
	the inflaton condensation is completely broken into particles and participates in the thermal plasma.
}
Throughout this paper, we assume the under occupied case, in
which the decay of the scalar field is described by the field theory in
the zero temperature.

As we have declared at the beginning of this section, let us consider
large angle scatterings.
In order to achieve thermalization, scatterings which increase the number
density of the particles are necessary.
The most efficient ones would be two body to three body scatterings. 
Cross sections of such scatterings, $\sigma_i$, are as large as
\begin{align}
 \sigma_i &\simeq k \frac{1}{p_i^2}\simeq k
  \frac{1}{m_\phi^2},\nonumber\\
 k &\simeq \frac{g^6}{128\pi^3},
\end{align}
where $g$ is a gauge or Yukawa coupling constant.
For example, QCD interactions yield $k\simeq 10^{-3}$.
Therefore, just after the end of reheating, an average
interaction rate is given by
\begin{align}
\Gamma_{\rm int}\simeq \vev{\sigma n v} \simeq \sigma_i n_i \simeq k\frac{\Gamma_\phi^2 \mpl^2}{m_\phi^3}.
\end{align}

A ratio between the interaction rate and the Hubble parameter just after
the end of reheating is given by
\begin{align}
 \frac{\vev{\sigma nv}}{H}\simeq k\frac{\Gamma_\phi \mpl^2}{m_\phi^3}.
\end{align}
This ratio is larger than one if
\begin{align}
 \Gamma_\phi > \frac{1}{k}\frac{m_\phi^3}{\mpl^2} \gtrsim \frac{1}{k}\times(
 \text{Decay width by Planck-suppressed interactions}).
\end{align}
In this case, thermalization occurs just after the end of
reheating solely by large angle scatterings.

Therefore, if the decay of the scalar field occurs via
interactions stronger than Planck-suppressed ones, 
the decay products of the scalar field thermalize
soon after the end of reheating. This is due to large number density
of the decay products.
On the other hand, if the decay takes place via Planck-suppressed interactions, thermalization does
not occur soon after the end of reheating, as long as large angle
scatterings are concerned.

What will happen if thermalization is controlled solely by large angle
scatterings,
 when the decay is via Planck-suppressed interactions?
As Universe expends, momenta of the decay products, number density, and
the Hubble parameter decrease in proportional to $a^{-1}$, $a^{-3}$, and
$a^{-2}$, where $a$ is the scale parameter of the Universe. Hence, the
relation between the interaction rates and the Hubble rate increases in
proportional to $a^1$.
Therefore, thermalization occurs when the Universe becomes $k^{-1}$
times larger after the end of the reheating process. The
thermalized temperature $T_{\rm th}$ is as large as
\begin{align}
 T_{\rm th}\simeq k T_\text{rh},
\end{align}
which is smaller than $T_\text{rh}$ even if the QCD interaction is involved.
If the scalar field does not decay into particles with the QCD interaction,
$T_{\rm th}$ would be far smaller than $T_\text{rh}$.

\section{Closer look at thermalization processes}
\label{sec:under}

In the previous section,  the simple analysis has revealed that large
angle scatterings alone cannot thermalize the particles produced by the
decay of the scalar field instantaneously for Planck-suppressed decay.
However, in this section, we will see that actual thermalization occurs much faster
with a closer look at thermalization processes, following 
the discussion in Ref.~\cite{Kurkela:2011ti}.
See also Ref.~\cite{Arnold:2002zm}.
This is because soft particles radiated from decay products with large momenta
play important roles in thermalization, 
which is neglected in the previous section.
Consequently, as we will see in Sec.~\ref{sec:application}, thermalization after reheating occurs rather instantaneously 
even for Planck-suppressed decay.

First, let us recall the discussion in Sec.~\ref{sec:review}.
We have shown that at each point during and after reheating, 
although the energy and number densities of decay products are dominated 
by particles with momenta $p\sim m_\phi$, their distribution becomes complicated 
due to the red-shift of previously produced particles: Eq.~\eqref{eq:dist_dr_reh_1}.
As a first step, it is instructive to 
neglect the expansion of the Universe and  consider the case 
in which the distribution function is given by a mono-spectrum one:
\begin{align}
	f (p) \simeq 
	\begin{cases}
		n_h / m_\phi^3 &\text{for}~~p \sim m_\phi \\
		0 &\text{for others}
	\end{cases},
\end{align}
where $n_h$ is the number density of the decay products with large momenta,
which we call the ``hard primaries''.
Effects of previously produced particles are separately discussed in Sec.~\ref{sec:dr_reh}.
There, we find that the effects of the previously produced particles are negligible 
in estimating the thermalization time scale and hence the mono-spectrum approximation is an appropriate description.

Throughout this paper, we assume that the decay products interact with each other by gauge interactions.
Following Ref.~\cite{Kurkela:2011ti}, the distribution is expressed by the fine
structure constant of a gauge interaction $\alpha\equiv g^2/\left(4\pi\right)$ as
\begin{align}
	\alpha^{-c} \equiv n_h / p_i^3,
\end{align}
where $p_i =m_\phi$.%
\footnote{The following discussion is a general one and the assumption that $p_i = m_\phi$  is not necessary.}
Cases with $c > 0$ are referred to as over occupied systems and those with $c < 0$ are referred to as under occupied systems,
and processes of thermalization are found to be different for each
cases~\cite{Kurkela:2011ti}.

As we have mentioned in the previous section,
we are particularly interested in the case with
\begin{align}
	\alpha^{-c} = \frac{\Gamma_\phi^2 \mpl^2}{m_\phi^4},
\end{align}
which typically implies $c<0$ for the Planck-suppressed decay.\footnote{
For instance, $\alpha^{-c} \sim 10^{-10} (m_\phi / 10^{13}\,\GEV)^2$ for the dimension 5 Planck-suppressed decay.
}
Hence, let us concentrate on the under occupied case defined as $c<0$ in the following.

\subsection{Basic ingredients}
\label{sec:basic}

Before reviewing thermalization in the under occupied case, let us
briefly summarize basic ingredients for the thermalization process~\cite{Arnold:2002zm,Kurkela:2011ti}.
Here, we consider thermalization by a gauge interaction, 
which is effective due to t-channel enhancement as we will see later.
In the medium, the dispersion relation of particle-like excitation, so-called quasi-particle, 
can be approximated with
\begin{align}
	\omega (\bm{p}) \simeq \sqrt{\bm{p}^2 + m_s^2};~~~m_s^2 \sim \alpha \int_{\bm{p}} \frac{f(p)}{p},
\end{align}
where $\int_{\bm{p}} \equiv \int d^3 p /(2\pi)^3$ , $m_s$ is the
screening mass, and $f(p)$ is the distribution function of the ``screener'',
which interacts with the quasi-particle.

The non-equilibrium dynamics of plasma can be studied by the Boltzmann equations of quasi-particles
with collision terms of spatially localized interactions which can be obtained from $S$-matrices,
if the system respects the following two conditions: 
(i).~the typical size of quasi-particles is smaller than the mean free path,
(ii).~the typical duration of interaction is shorter than the mean free time.
Note that when the Landau Pomeranchuk Migdal (LPM) effect~\cite{Landau:1953um,Arnold:2001ba,Arnold:2002ja,Besak:2010fb} becomes efficient, 
the condition (ii) does not hold and hence we need special cares
as we will see later.
If both (i) and (ii) are met, the distribution function $f(t, \bm{p})$ obeys the Boltzmann equation:
\begin{align}
	\left[ \der_t - H \bm{p} \cdot \frac{\der}{\der \bm{p}} \right] f(t,p) = - {\cal C} [f],
\end{align}
where $H$ is the Hubble parameter and ${\cal C}$ represents a collision term via spatially
local interactions. Here we gave restricted ourselves to an isotropic system which we are interested in.
The collision term $\cal{C}$ imprints basic elementary processes: number conserving two-to-two scattering and
number violating nearly collinear in-medium splitting.

\begin{figure}[tb]
 \begin{center}
     \includegraphics[width=0.4\linewidth]{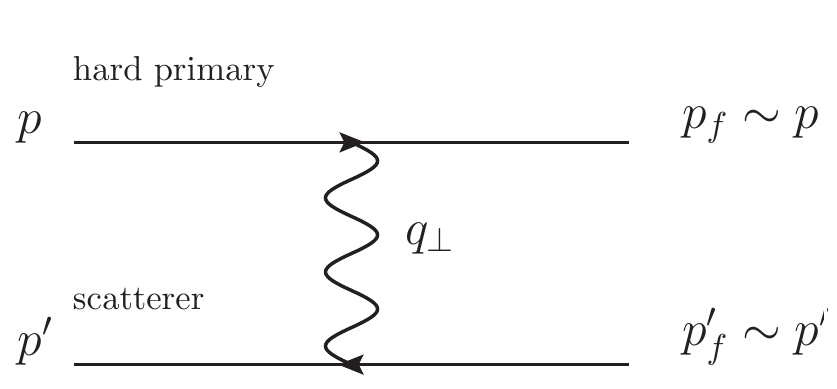}
 \end{center}
\caption{Two-to-two scattering.}
\label{fig:two-two}
\end{figure}

The former process (Figure.~\ref{fig:two-two}) diffuses the momentum of particles and
the average of squared momentum transfer
is described by a random walk:
\begin{align}
	\left( \Delta p \right)^2 \sim \hat q_\text{el} t,
\end{align}
with $\hat q_\text{el}$ being a diffusion constant that is estimated
as
\begin{align}
	\hat q_\text{el} \sim \int d^2 q_\perp  \frac{d \Gamma_\text{el}}{d q_\perp^2} q^2_\perp
	\sim \alpha^2 \int_{\bm{p'}} f(p') \left[ 1 \pm f(p') \right],
\end{align}
where $f(p')$ is the distribution function of ``scatterers'', which
scatter the hard primaries off. The sign $+/-$ corresponds to the case
in which the scatterers are bosonic/fermionic.
Here $\Gamma_\text{el}$ is an elastic scattering rate with a 
momentum exchange $q_\perp <p$, where $q_\perp$ is a momentum transfer.

For example, the scattering by the t-channel exchange of the gauge boson yields
\begin{align}
\label{eq:soft elastic}
	\frac{d \Gamma_\text{el}}{d q_\perp^2} \sim \frac{\alpha^2}{q_\perp^2 \left( q_\perp^2 + m_s^2 \right)} 
	\int_{\bm{p'}} f(p') \left[ 1 \pm f(p') \right].
\end{align}
Due to enhancement for small $q_\perp$, this scattering, if it exists, plays important roles.%
\footnote{
This enhancement is absent for the t-channel exchange of a scalar
particle by a Yukawa interaction.
}
In fact, the t-channel enhanced soft elastic scatterings occur so frequently
that the cosmic expansion can be neglected:
$H \ll \Gamma_\text{el}^\text{(soft)} \sim \hat q_\text{el} / m_s^2$.
Therefore, we neglect the cosmic expansion at first
and compare the time scale of bottleneck process with that of cosmic expansion finally.

\begin{figure}[tb]
 \begin{center}
  \includegraphics[width=0.4\linewidth]{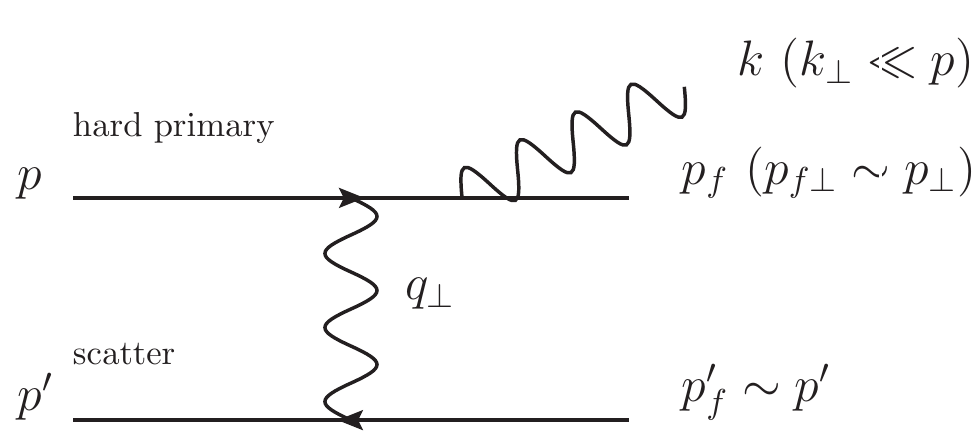}
 \end{center}
\caption{One example of collinear in-medium splittings without the LPM effect.
}
\label{fig:one-two}
\end{figure}

On the other hand, since the soft elastic scattering with $q_\perp \sim m_s$
makes the quasi-particle slightly off-shell in the medium (in-medium width),
the latter process
change the number by radiating a nearly collinear gauge boson (Figure~\ref{fig:one-two}).
As mentioned earlier, if each soft elastic scatterings occur more frequently than
the duration of the splitting process,
the condition (ii) is violated and one has to take the quantum mechanical interference of each scatterings into account,
that is, we have to care about the LPM effect~\cite{Landau:1953um,Arnold:2001ba,Arnold:2002ja,Besak:2010fb}.

Let us estimate the time scale of quantum interference.
Suppose that the parent particle travels
through a spatial distance $\bm{x} = t  \bm{\hat x}$ with a time interval $t$ [$x = (t,\bm{x})$] 
under soft elastic scatterings.
The interference of bremsstrahlung produced by the first and last scatterings, at $\bm{0}$ and $\bm{x}$,
remains until the phase factor varies significantly,
$1 \gtrsim k \cdot x \sim t k \theta^2$,
with $k$ being the momentum of the daughter particle and $\theta$ being the emission angle of the daughter particle.

When the daughter particle is a non-abelian gauge field, 
it is also charged and
is subjected to the soft elastic scatterings.
The coherence is lost if the daughter particle changes its direction substantially due to these random soft collisions,
$1/\sqrt{kt} \sim \theta (\sim k_\perp/k)$.
Recalling that the perpendicular momentum diffuses via elastic scatterings as $k_\perp^2 \sim \hat q _\text{el} t$,
one finds the quantum mechanical formation time scale as\footnote{
	Note that at least one soft elastic scattering is required for the collinear splitting to occur.
}
\begin{align}
	t_\text{form}^\text{NA} \sim \max \left[ 1/\Gamma_\text{el}^{(\text{soft})}, \sqrt{k/\hat q_\text{el}} \right]
\end{align}
where $\Gamma_\text{el}^{(\text{soft})}$ is the rate of one soft elastic scattering $q_\perp \sim m_s$,
which is given by $\Gamma_\text{el}^{(\text{soft})} \sim \hat q_\text{el}/m_s^2$.
Then, the rate of the collinear splitting in the medium is roughly estimated as
\begin{align}
	\Gamma_\text{split}^\text{NA}(k) \sim \alpha t_\text{form}^{\text{NA}^{-1}}(k)
	\label{eq:split}
\end{align}
with $k$ being the momentum of the daughter particle.
Note that the LPM effect becomes efficient when the momentum $k$ exceeds a typical value:
\begin{align}
\label{eq:klpm_NA}
k_\text{LPM}^\text{NA} \equiv m_s^4 / \hat q_\text{el}. 
\end{align}

When the daughter particle is an abelian gauge field,
however, it is not charged under its own gauge group.
Therefore, the coherence remains unless the parent particle changes its direction substantially,
$\theta\sim p_\perp / p\sim1/ \sqrt{kt}$ with $p$ being the momentum of the parent particle.
Contrary to the non-abelian case, the formation time scale depends on the momentum of the parent particle as
\begin{align}
	t_\text{form}^\text{U(1)} \sim \max \left[ 1/\Gamma_\text{el}^{(\text{soft})}, \sqrt{p^2/ (k \hat q_\text{el})} \right].
\end{align}
Note that the LPM effect dominates the collinear splitting when the momentum k becomes smaller than 
a typical value:
\begin{align}
\label{eq:klpm_abelian}
 k_\text{LPM}^\text{U(1)} \equiv\hat q_\text{el} Q^2 / m_s^4. 
\end{align}

\subsection{Thermalization with non-abelian charged primaries}
\label{sec:th_na}

We are now in a position to discuss how thermalization proceeds.
In the following, we consider the case in which the majority of primary particles produced
from the decay of the scalar field is
charged under some non-abelian gauge group or non-abelian gauge bosons
themselves.
We consider that this case is practically most important.%
\footnote{
For example, as long as the decay of the scalar field respects 
the grand unified theory, the majority of the decay products always has non-abelian gauge interactions.
The case with an abelian gauge interaction is discussed elsewhere~\cite{khkm}. There, we find that thermalization 
proceeds slower than the case of the non-abelian gauge interaction.
This is because the gauge boson, which plays an important role in the thermalization process as we will see, is not 
charged under its own gauge group in abelian gauge theories.
}
The thermalization of QCD is discussed in
Ref.~\cite{Kurkela:2011ti} and we closely follow the discussion there.
As we will see,
soft particles produced by collinear splittings via small angle scatterings play crucial roles
in thermalization after/during reheating.

\begin{figure}[tb]
 \begin{center}
  \includegraphics[width=0.6\linewidth]{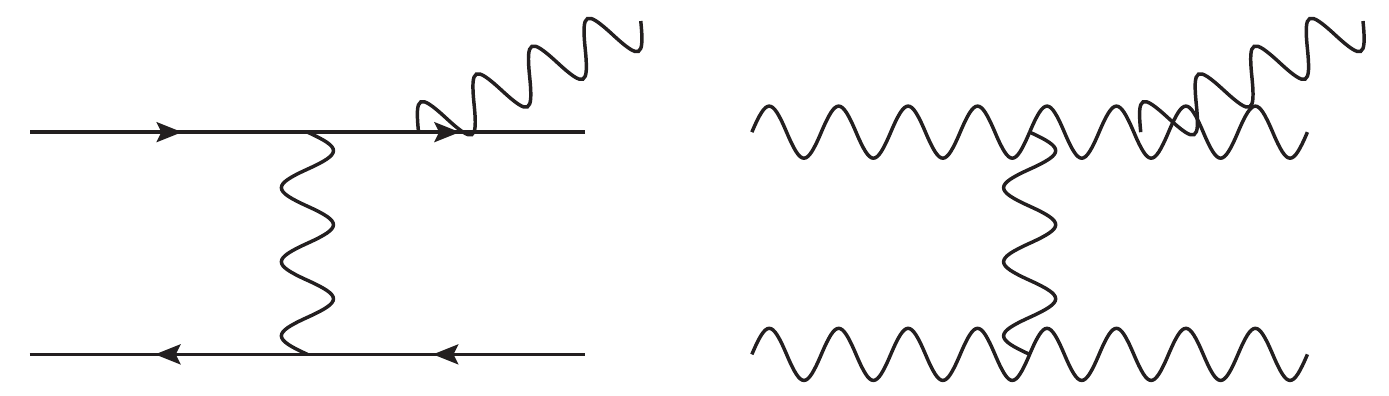}
 \end{center}
\caption{Examples of creation of soft population by collinear splittings for the
 non-Abelian case without the LPM effect.
}
\label{fig:nonabelian_split}
\end{figure}

First of all, soft bosons are radiated from the primary hard particles with the momentum $p_i \sim m_\phi$
via collinear splittings in association with t-channel enhanced
soft elastic scatterings, whose rate is given in Eq.~(\ref{eq:soft elastic}),
and a new population around small momentum is created (Figure~\ref{fig:nonabelian_split}).
The phase space distribution of these soft particles is estimated as
\begin{align}
	f_s (t,p) \sim \Gamma_\text{split}^\text{NA} (p) n_h p^{-3} t
	\sim \begin{cases}
		\alpha \Gamma_\text{el}^{(\text{soft})} n_h p^{-3} t &\text{for}~~p < k_\text{LPM}^\text{NA} \\
		\alpha \sqrt{\hat q_\text{el}} n_h p^{-7/2} t &\text{for}~~ k_\text{LPM}^\text{NA} < p
	\end{cases}. \label{eq:dist_soft}
\end{align}
For later convenience, let us summarize quantities determined by the
number density of the hard primaries:
\begin{align}
 m_s^2 \sim \alpha^{1-c}p_i^2,~~\hat{q}_\text{el}\sim \alpha^{2-c}p_i^3,~~k^\text{NA}_\text{LPM}\sim\alpha^{-c}p_i \left( < m_\phi \right).
\end{align}

Note that the spectrum is quite sharp compared with a thermal-like distribution, that is proportional to $p^{-1}$
and hence the ``soft sector'' is effectively over occupied.
They interact among themselves and with hard particles, 
and fall into a thermal-like distribution below a scale $p_\text{max}$:
\begin{align}
	f_s \sim T_\ast/p;~~\text{for}~~p < p_\text{max}.
\end{align}

The typical scale $p_\text{max}$ evolves towards a UV regime
via interactions not only among soft particles but with the remaining hard particles.
Let us estimate the typical time scale of $p_\text{max}$ evolution.
In the following, let us consider the case with $m_s < k_\text{LPM}$, that is $c > -1$.
For $c < -1$, the screening mass is larger than $k_\text{LPM}$ from the beginning
and the production of soft particles is always LPM suppressed.
In this case, one can skip the following discussion until Eq.~\eqref{eq:eff temp}.
The important processes that drive $p_\text{max}$ towards UV are elastic scatterings with hard particles and
saturation between the production and absorption of soft particles.
The former process leads to the following growth rate of $p_\text{max}$:
\begin{align}
	p_\text{max}^{(\text{h scat})} \sim \sqrt{\hat q_\text{el}^{(\text{hard})} t} \sim \alpha^{1- c/2} p_i \sqrt{p_i t }.
	\label{eq:hard_scat}
\end{align}
The latter process is characterized by the balance of production and absorption rate for the soft population:
\begin{align}
	0 &= \int_{\bm {p}} \Gamma_\text{split} \left[ f (p) \left( 1 + f (p-k)  \right)  \left( 1 + f_s (k)  \right)
	- f (p-k) f_s (k) \left( 1 + f (p) \right) \right] \\
	& \sim \int_{\bm{p}} \Gamma_\text{split} 
	\left[ k f' (p) f_s (k) + f(p) \right],
\end{align}
which implies
\begin{align}
	f_s (k) \sim \frac{\int_{\bm{p}} f(p)}{\int_{\bm{p}} - k f'(p)} \sim \frac{p_i}{k}.
\end{align}
Hence, if the distribution of soft modes exceeds $f_s (k) \sim p_i/k$,
then the production of soft particles is cancelled by the absorption.
By equating this condition with Eq.~\eqref{eq:dist_soft}, one finds the typical scale below which
the distribution turns into the thermal-like one:
\begin{align}
	p_\text{max}^{(\text{sat})} \sim \alpha^{1- c/2} p_i \sqrt{p_i t }
	\label{eq:sat},
\end{align}
that is equal to Eq.~\eqref{eq:hard_scat}.
Note that the effective temperature for the soft sector is given by
\begin{align}
	T_\ast \sim p_i,
\end{align}
that remains constant at this regime.
Also note that still, the screening mass, the energy and number densities are dominated by the remaining hard particles.

As the the scale $p_\text{max}$ evolves, the screening mass becomes dominated by the soft particles and
also the LPM scale $k_\text{LPM}^\text{NA}$ becomes comparable to $p_\text{max}$.
This time scale is estimated as
\begin{align}
\label{eq:T*evolution}
	\left( p_i t \right)_\text{LPM} \sim \alpha^{-2-c}.
\end{align}
After that, the production rate of soft particles given in Eq.~\eqref{eq:split} is dominated by the LPM suppressed one,
and hence the effective temperature no longer remains constant:%
\footnote{
Here, we have assumed that $\hat{q}_\text{el}$ is determined by the hard
primaries.
From Eq.~(\ref{eq:T*evolution}), it can be shown that the contribution of the
soft sector to $\hat{q}_\text{el}$ is as large as that of the hard primaries.
}
\begin{align}
	T_\ast \sim p_\text{max} \left( \alpha \sqrt{\hat q_\text{el} / p_\text{max}} \right) n_h t / p_\text{max}^3
	\sim \alpha^{- c/4 - 1/2} p_i \left( p_i t \right)^{-1/4}.
	\label{eq:eff temp}
\end{align}
The soft sector ``thermalizes'' when the effective temperature $T_\ast$ becomes comparable to $p_\text{max}$,
which occurs at
\begin{align}
	 \left( p_i t \right)_\text{soft th} \sim \alpha^{ - 2 + c/3}.
	 \label{eq:soft_th}
\end{align}
Around that time, not only the screening mass but the number density are dominated by soft particles:
$n_s \sim T_\ast^3 \gtrsim n_h \sim \alpha^{-c} p_i^3$.
However, the energy density is still dominated by the remaining hard particles.

Finally, the remaining hard particles dissipate their energy to the soft ``thermal'' sector
via multiple splitting of daughter particles.
One can determine the maximum scale $k_\text{split}$ below which a particle can deposit an order one fraction of
its energy into the background soft ``thermal'' sector from a criteria 
$\Gamma^\text{NA}_\text{LPM} (k_\text{split}) t \sim 1$.
The remaining hard particles lose their energy dominantly via the splitting of ``hardest'' daughter particle
with momentum $k_\text{split}$.
Through this process, the ``thermal'' sector is heated, and the energy conservation implies
$T_\ast^4 \sim k_\text{split} n_h$.
These equations suggest
\begin{align}
	T_\ast &\sim \alpha^{4-c} p_i (p_i t)^2, \\
	k_\text{split} &\sim \alpha^{-3c + 16} p_i \left( p_i t \right)^{8}.
\end{align}

Note that time scales of processes
which establish the chemical equilibrium for the soft sector is as large as
\begin{align}
(\alpha^2 T_*)^{-1} \sim \alpha^{-6+c} p_i^{-1} (p_i t)^{-2},
\end{align}
and shorter than the time scale $t$ as long as $t>t_\text{soft th}$.
Therefore, soon after hard particles emit daughter particles with momenta $k_\text{split}$ and the daughter particles deposit their energy into the soft sector, particles in the soft sector follow the chemical equilibrium.
The remaining hard particles completely lose their energy when this scale becomes comparable to $p_i$.
As a result, thermalization is completed at
\begin{align}
	\left( p_i t \right)_\text{split} \sim \alpha^{- 2 + 3c / 8 },
\end{align}
and the thermal bath with the temperature $T_\text{rh}$ is created.

Since $c<0$,
the relevant time scales satisfy the following inequality:
\begin{align}
	\left( p_i t \right)_\text{LPM} < \left( p_i t \right)_\text{soft th} < \left( p_i t \right)_\text{split}.
\end{align}
Thus, the bottleneck process is one in which the remaining hard
particles lose their energy into the background plasma,
\footnote{
The time scale of this final process is parametrically the same as the dissipation time scale
for the hard particle with momentum $p_i > T_\text{rh}$ to travel through the thermal plasma with the temperature $T_\text{rh}$:
\begin{align}
	t_\text{diss} \sim \left( \alpha^2 T_\text{rh} \right)^{-1} \sqrt{p_i / T_\text{rh}} \sim t_\text{split}.
\end{align}
}
and hence the timescale of thermalization $t_\text{th}$ is given by~\cite{Kurkela:2011ti}
\begin{align}
\label{eq:tth_NA}
 t_\text{th} \sim \frac{1}{p_i}\alpha^{-2 + 3c/8}.
\end{align}

\begin{figure}[tb]
 \begin{center}
  \includegraphics[width=0.4\linewidth]{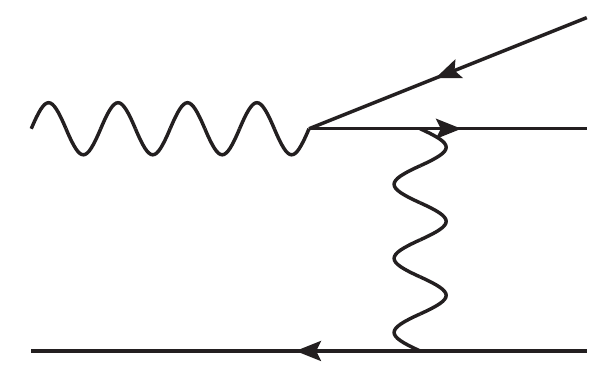}
 \end{center}
\caption{Example of creation of soft population by collinear splittings for the
 Abelian case without the LPM effect.
}
\label{fig:abeliansplit}
\end{figure}

Before closing this section, let us comment on abelian gauge bosons.
Decay products of the scalar field often include abelian gauge bosons.
Even if it is not the case, decay products in general have an abelian gauge interaction
and hard primaries radiate not only non-abelian gauge bosons but also abelian gauge bosons. 
These hard abelian gauge bosons can dissipate their energies via collinear splittings of matters
with t-channel enhancement (Fig.~\ref{fig:abeliansplit}).
Therefore, they do not affect the above discussion.

\section{Application to reheating process}
\label{sec:application}
In this section, let us study the implication of the result given in
Eqs.~(\ref{eq:tth_NA}) to the reheating process.
We discuss the thermalization of decay products after and during
reheating separately.

\subsection{After reheating}
\label{sec:}

At first,
it is instructive to consider an instantaneous decay approximation for the scalar field.
In this case, the spectrum of hard primaries is approximated with a mono-spectrum one
and we can neglect effects of red-shifted previously produced particles.
Actually, in the next section, we show that the effects of previously produced particles is negligible.
Let us summarize the conditions where the produced particles thermalize much faster than
the cosmic time scale.

Thermalization is completed against the expansion of the Universe if
the times scale of the thermalization is much shorter
than the inverse of Hubble parameter:
\begin{align}
\label{eq:condition_after_NA}
	1 \gg H t_\text{split} \sim \alpha^{-2} \left( \frac{m_\phi}{\mpl} \right)^{3/4} \left( \frac{\Gamma_\phi}{m_\phi} \right)^{1/4}.
\end{align}
Interestingly, this inequality implies that
thermalization is more likely to be completed for a smaller decay rate.
Thus,
in order to show that thermalization occurs instantaneously in general,
it is sufficient to discuss the largest decay rate.
Since we assume that the decay of the scalar field is well-described by
the field theory in the zero temperature, the decay rate should be
smaller than $m_\phi^2/\mpl$.
In this case, Eq.~(\ref{eq:condition_after_NA}) reads
\begin{align}
 \alpha \gg \frac{m_{\phi}^{1/2}}{\mpl^{1/2}} = 0.002\left(\frac{m_{\phi}}{10^{13}~\text{GeV}}\right)^{1/2}.
\end{align}
Note that the condition is severer for larger $m_{\phi}$.
The largest mass of the scalar field which appears in the cosmology would be the mass of an inflaton which has a convex potential.
For example, in chaotic or hybrid inflation models with convex potentials, the mass is as large as 
$10^{15}$ GeV~\cite{Dimopoulos:1997fv,Takahashi:2010ky,Harigaya:2012pg}.
One can easily see that unless the gauge coupling is extremely suppressed, the
decay products thermalize rather instantaneously
compared with the cosmic time scale.
As we have mentioned in the introduction,
this condition is much milder than that obtained in Ref.~\cite{Davidson:2000er},
which is given by $\alpha^3 \gg (m_\phi / \mpl)$,
since the growth of the soft sector promotes the energy dissipation of hard particles.

\subsection{During reheating}
\label{sec:dr_reh}

Then, let us discuss the effects of previously produced particles
while following the evolution of the momentum distribution of radiation during reheating.
As mentioned at the end of Sec.~\ref{sec:review},
the scalar field decays gradually before the completion of reheating.
Therefore previously produced particles are red-shifted, which results in the following distribution of hard particles [Eq.\eqref{eq:dist_dr_reh_1}]:
\begin{align}
	f_h (p) \sim 
	\begin{cases}
		\left( \cfrac{\Gamma_\phi \mpl^2}{m_\phi^3} \right) \left( m_\phi t \right)^{-1}
		\left( \cfrac{m_\phi}{p} \right)^{3/2} &\text{for}~~(t/t_i)^{-2/3} m_\phi \lesssim p \lesssim m_\phi \\[10pt]
		0 &\text{for~~otherwise}
	\end{cases}.
\end{align}
Since the number density is dominated by the hardest particles with $p \sim m_\phi$,
the soft particles are produced dominantly from them.
Hence, we can estimate the production of soft particles in the same way as Sec.~\ref{sec:th_na}.

To avoid possible complications,
we focus on the situation where the produced soft particles have enough time to thermalize against the expansion of the Universe.
In order to estimate the time after which the soft sector is expected to thermalize immediately,
let us recall the time scale $t_\text{soft th}$ which is given by Eq.~\eqref{eq:soft_th}:
\begin{align}
	\left(m_\phi t \right)_\text{soft th} \sim \alpha^{-2 + c/3}
	\leftrightarrow 	
	\left(m_\phi t \right)_\text{soft th} \sim 
	\alpha^{-3} \left( \frac{m_\phi^3}{\Gamma_\phi \mpl^2} \right)^{1/2}.
\end{align}
The soft sector thermalizes within the cosmic time scale for
\begin{align}
	1 > t_\text{soft th} H \leftrightarrow t > t_\text{soft th}.
\end{align}
As discussed previously, at that time, the number density is dominated by the soft sector
while the energy density is still dominantly stored in the hard sector.
Note that  the production of soft particles is already dominated by the LPM suppressed one.

The hard particles dissipate their energy into the soft thermal plasma via multiple splittings.
As done in the previous section,
we can estimate the splitting momentum $k_\text{split}$ below which a particle can deposit an order one fraction of its energy and
participate in the thermal plasma, from the criteria: $\Gamma^\text{NA}_\text{LPM} (k_\text{split}) t \sim 1$.
The thermal plasma is heated by the splitting of hardest particle together with the daughter particle 
of the momentum $k_\text{split}$,
since the hardest particles dominate the number density of hard sector.
Hence
the energy conservation gives $T_\ast^4 \sim n_h k_\text{split}$ with $T_\ast$ being the effective temperature of
soft thermal plasma.
From these equations, one finds
\begin{align}
	T_\ast &\sim \alpha^{4} \left( \frac{\Gamma_\phi \mpl^2}{m_\phi^3} \right) m_\phi \left( m_\phi t \right), \\
	k_\text{split} &\sim \alpha^{16} \left( \frac{\Gamma_\phi \mpl^2}{m_\phi^3} \right)^{3} m_\phi 
	\left( m_\phi t \right)^{5}.
\end{align}

Therefore, the distribution function of soft particles for $t > t_\text{soft th}$ can be roughly parametrized as
\begin{align}
	f_s(t,p)& \sim \begin{cases}
		\left( \cfrac{T_\ast}{p} \right)\cdots\,\text{boson}~~\text{or}~~1\cdots\,\text{fermion}~~ &\text{for}~~ p < T_\ast \\[15pt]
		\exp \left( - \cfrac{p}{T_\ast} \right) &\text{for}~~ T_\ast < p < k_\text{split}	\\[15pt]
		\alpha^{8} \left( \cfrac{\Gamma_\phi \mpl^2}{m_\phi^3} \right)^{5/2} \left( m_\phi t \right)^{3/2} 
		\left( \cfrac{m_\phi}{p} \right)^{7/2} &\text{for}~~  k_\text{split}< p < m_\phi
		\end{cases},
\end{align}
and that of hard primaries is given by
\begin{align}
	f_h(t,p)& \sim \begin{cases}
		\left( \cfrac{\Gamma_\phi \mpl^2}{m_\phi^3} \right) \left( m_\phi t \right)^{-1} 
		\left( \cfrac{m_\phi}{p} \right)^{3/2} &\text{for}~~(t/t_i)^{-2/3} m_\phi  < p < m_\phi \\[15pt]
		0 &\text{for}~~ m_\phi < p.
	\end{cases}.
\end{align}
Note that the soft sector already dominates over the red-shifted hard particles below an intermediate scale
\begin{align}
	k_\text{int} \sim \alpha^4 \left( \frac{\Gamma_\phi \mpl^2}{m_\phi^3} \right)^{3/4}
	m_\phi \left( m_\phi t \right)^{5/4},
\end{align}
which is larger than $k_\text{split}$ for $t<t_\text{split}$ (see below)
and hence the effects of red-shifted particles are not so relevant.

The remaining hard particles completely lose their energy when the splitting momentum becomes comparable to the maximum momentum:
$k_\text{split} \sim m_\phi$.
Consequently, thermalization is completed at
\begin{align}
	\left( m_\phi t \right)_\text{split} \sim \alpha^{-16/5} 
	\left( \frac{\Gamma_\phi \mpl^2}{m_\phi^3}  \right)^{-3/5}.
\end{align}
At that time, the intermediate scale also reaches the maximum momentum simultaneously: $k_\text{int}|_{t_\text{split}} \sim m_\phi$.
The temperature of the thermal bath at that time, which we refer to as the maximum temperature $T_\text{max}$, is given by
\begin{align}
T_\text{max} \sim \alpha^{4/5} m_\phi \left( \frac{\Gamma_\phi \mpl^2}{m_\phi^3}\right)^{2/5}.
\end{align}

After $t > t_\text{split}$,
particles produced from the decay of the scalar field
thermalize instantaneously,
and eventually reheating is completed at $t_\text{rh} \sim \Gamma_\phi^{-1}$.
Note that the condition for the thermalization before reheating is completed
is nothing but $t_\text{split} < t_\text{rh}$, which results in Eq.~\eqref{eq:condition_after_NA}.
Relevant time scales are summarized in Tab.~\ref{tab:time}.

\begin{table}[H]
\vskip-0.5em
\caption{Relevant time scales.}
\begin{center}
{\renewcommand\arraystretch{1.5}
\begin{tabular}{lccc}
\hline
\hline
time scale & general && dim 5 \\
\hline
$\left( m_\phi t \right)_\text{soft th}$ & $\alpha^{-3} \left( \cfrac{\Gamma_\phi \mpl^2}{m_\phi^3} \right)^{-1/2}$ &&
$\alpha^{-3}$ \\[10pt]
$\left( m_\phi t \right)_\text{split}$ &$\alpha^{-16/5} \left( \cfrac{\Gamma_\phi \mpl^2}{m_\phi^3} \right)^{-3/5}$ && $\alpha^{-16/5}$ \\[10pt]
$\left( m_\phi t \right)_\text{rh}$ & $\left( \cfrac{\Gamma_\phi}{m_\phi} \right)^{-1}$ && $\cfrac{\mpl^2}{m_\phi^2}$ \\
\hline
\hline
\end{tabular}
}
\end{center}
\label{tab:time}
\end{table}%

\section{Conclusion and Discussions}
\label{sec:conc}

In this paper, we have investigated the dynamics of reheating and
thermalization of light particles which are produced from the perturbative decay of the scalar field,
especially
paying attention to the case with a small decay width of the scalar field, such as the one given by
Planck-suppressed interactions.
Since the number density is under occupied in this case, 
number violating splittings with small momentum exchanges and
the evolution of the soft sector play crucial roles in the dynamics towards thermal equilibrium.
Importantly,
the bottleneck process is the dissipation of remaining hard particles into the soft sector as pointed out 
in Refs.~\cite{Baier:2000sb,Kurkela:2011ti}
in the context of QGP.

Following their arguments,
we have studied thermalization after/during reheating in detail, while paying attention to the red-shift of 
previously produced particles before the completion of reheating and
to the dilution of their number density by the cosmic expansion.
We have shown that thermalization takes place instantaneously in most cases,
and also found that red-shifted previously produced particles are not relevant
since radiated soft particles dominate the number density of red-shifted previously produced ones
and that radiation is always dominated by that from hardest primaries which are not red-shifted.
The condition under which produced particles thermalize
before the completion of reheating
is given by
\begin{align}
	\alpha^{8/5} \gg \frac{m_\phi}{\mpl} \left(\frac{\Gamma_\phi \mpl^2}{m_\phi^3}\right)^{1/5},
\end{align}
which turns out to be much milder than that obtained in the previous work~\cite{Davidson:2000er}
because the growth of the soft sector promotes the dissipation of remaining hard particles.
As a result, even if the mass of the scalar field is as large as $10^{15}\, \GEV$,
it is shown that thermalization takes place instantaneously in most cases.
Our result is applicable to general situations where a decay of a non-relativistic matter 
reheats the universe via a small decay width and the decay products contain particles that are charged under a non-abelian gauge group.

Our analysis can also be applied to the instant preheating scenario~\cite{Felder:1998vq} in which
produced light particles have small number densities compared with thermal ones.
The situation is quite similar
to the case we have considered
and hence our result is helpful in estimating the time scale
when non-perturbative particle production is shut off by rescatterings~\cite{Kofman:1994rk}
(See also \cite{Mukaida:2012qn}).

\section*{Acknowledgment}

The authors thank Shigeki Matsumoto  and Kazunori Nakayama for useful discussions.
This work is supported by the World Premier International Research Center Initiative (WPI Initiative), MEXT, Japan (K.H.), and JSPS Research Fellowships
for Young Scientists (K.H. and K.M.).

\appendix



\end{document}